\documentclass[11pt]{article}
\usepackage{moriond,epsfig}

\def\be{\begin{equation}}
\def\ee{\end{equation}}
\def\bea{\begin{eqnarray}}
\def\eea{\end{eqnarray}}

\newcommand{\pt}{\ensuremath{p_{\mathrm{t}}} }
\newcommand{\dNdy}{\mathrm{d}N_\mathrm{ch}/\mathrm{d}y }
\newcommand{\dNdeta}{\mathrm{d}N_\mathrm{ch}/\mathrm{d}\eta }

\begin{document}
\vspace*{4cm}
\title{PHYSICS OF THE ALICE EXPERIMENT}

\author{ I. BELIKOV \footnote{On leave from JINR, Dubna, Russia} for the
  ALICE Collaboration }

\address{Physics Department, CERN, CH-1211 Geneva 23,\\
Switzerland}

\maketitle\abstracts{
 A short description of the ALICE detector at CERN is given.
 The experiment is aiming to study the properties of the quark-gluon plasma
 by means of a whole set of probes that can be subdivided into three  
 classes: soft, heavy-flavour and high-\pt probes. Each of the classes
 is illustrated by a few typical examples.}

\section{Introduction}
A Large Ion Collider Experiment (ALICE)~\cite{Carminati:2004fp} at CERN 
is a general-purpose heavy-ion experiment designed to study the 
physics of strongly interacting 
matter and the Quark-Gluon Plasma (QGP) in nucleus-nucleus collisions at
the LHC. In addition
to heavy systems, the ALICE Collaboration will study collisions of lower-mass
ions, which are means of varying the energy density, and protons (both pp and
pA), which primarily provide reference data for the nucleus--nucleus
collisions. The pp data will also allow for a number of genuine pp physics 
studies.

\begin{figure}
\centering
\epsfig{figure=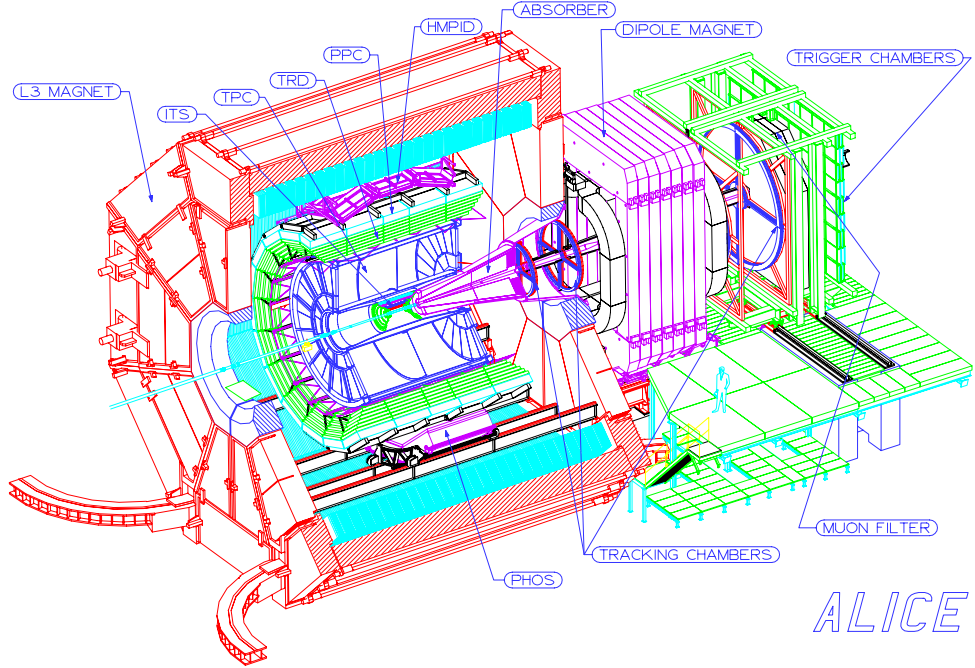,width=0.8\textwidth}
\caption{Schematic layout of the ALICE detector.
\label{fig:layout}}
\end{figure}

The detector consists of a central part (see Fig.~\ref{fig:layout}), which measures event-by-event hadrons,
electrons and photons, and of a forward spectrometer to measure muons.
The central part, which covers polar angles from $45^{\circ}$ to $135^{\circ}$
over the full azimuth, is embedded in the large L3 solenoidal magnet.
It consists of: an Inner Tracking System (ITS) of high-resolution
silicon detectors; a cylindrical Time-Projection Chamber (TPC);
three particle identification arrays of: Time-Of-Flight (TOF) detector, 
 Transition-Radiation Detector (TRD) and a single-arm ring imaging 
Cherenkov (HMPID); and a single-arm
electromagnetic calorimeter (PHOS).
The forward muon spectrometer (covering polar angles $180^{\circ} - \theta =
2^{\circ}$--$9^{\circ}$) consists of a complex arrangement
of absorbers, a large dipole magnet, and fourteen planes of tracking and
triggering chambers. Several smaller detectors for
global event characterization and triggering are located at forward angles.

The detector is optimized for charged-particle density $\dNdy =
4000$ and its performance is checked in detailed simulations up to 
$\dNdy = 8000$. The track reconstruction efficiency in the acceptance of
TPC is about 80\% down to $\pt\sim 0.2$~GeV/$c$ and about 90\% for tracks
with $\pt > 1$~GeV/$c$. It is limited only by the particle decays and small 
dead zones between the TPC sectors. Typical momentum resolution obtained
with the magnetic field of 0.5~T is $\sim 1$\% at $\pt\sim 1$~GeV/$c$ and
$\sim 4$\% at $\pt\sim 100$~GeV/$c$. The secondary vertices can be
reconstructed with the precision better then 100~$\mu$m.

    The detector has excellent PID capabilities. From $p\sim 0.1$~GeV/$c$
to a few GeV/$c$ the charged particles are identified by combining
the PID information provided by ITS, TPC, TRD, TOF and HMPID. Statistically,
the charged particles can be identified up to a few tens GeV/$c$ using
the relativistic rise of d$E$/d$x$ in TPC.  Electrons above 1~GeV/$c$
are identified by TRD, and muons are registered by the muon spectrometer.


\section{Probing the QGP}
The properties of the QGP state can be studied by means of many
observables.  In ALICE, the QGP observables (probes) are traditionally 
subdivided into three classes: soft probes (with the typical momenta
$p \le 1$~GeV/$c$), heavy-flavour probes (i.e. using the particles having 
c- and b-quarks) and high-\pt probes (in the momentum range above several
tens GeV/$c$).  Below we give a few examples of the planned observations 
belonging to each of these classes.
  
   \subsection{Soft QGP probes}
ALICE will measure the charged-particle multiplicity and the 
charged-particle pseudo-rapidity distribution over almost 8 units of $\eta$
by means of Forward Multiplicity Detector (FMD) and the innermost layers of
ITS. The way the initial energy is redistributed over the particles in the 
final state is strictly linked with such an important thermodynamical
quantity as the energy density $\varepsilon$ reached in the early phase of
the collision (see for example~\cite{Bjorken:1982qr}).

    Furthermore, the multiplicity information allows one to constrain the
hadroproduction models. Thus, in the phenomenological 
approach~\cite{Kharzeev:2000ph}, the measured pseudorapidity density 
$\dNdeta$ is 
expressed as the sum of a term proportional to the number of participants 
$N_{\rm part}$ (soft component) and the number of binary collisions 
$N_{\rm coll}$ (hard component). Measuring $\dNdeta$ as a function of 
$N_{\rm part}$, one can estimate the relative number of particles produced
in hard and soft scatterings.     

\begin{figure}
\centering
\epsfig{figure=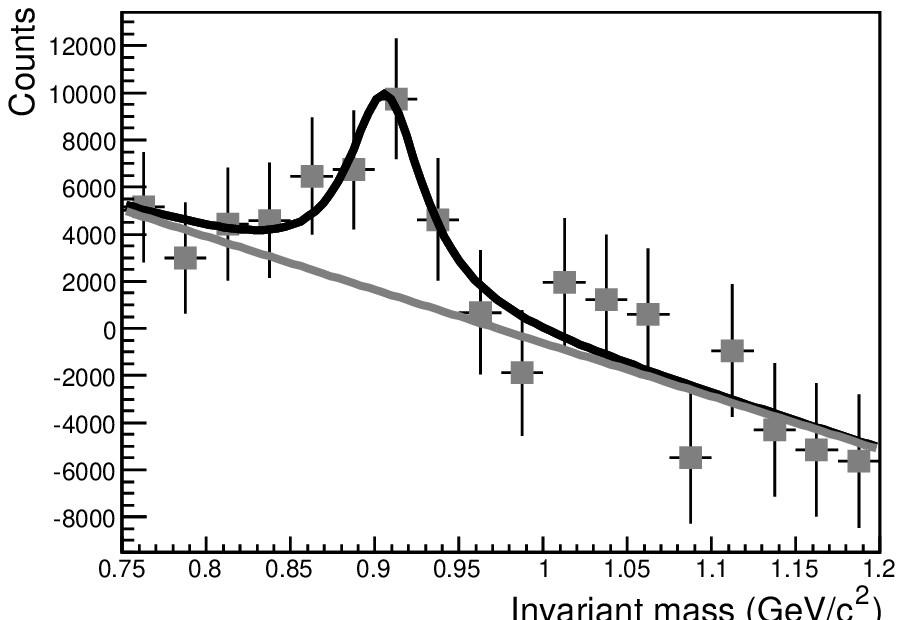,width=0.43\textwidth}
\includegraphics[width=0.55\textwidth, bb=0 0 540 300,clip]{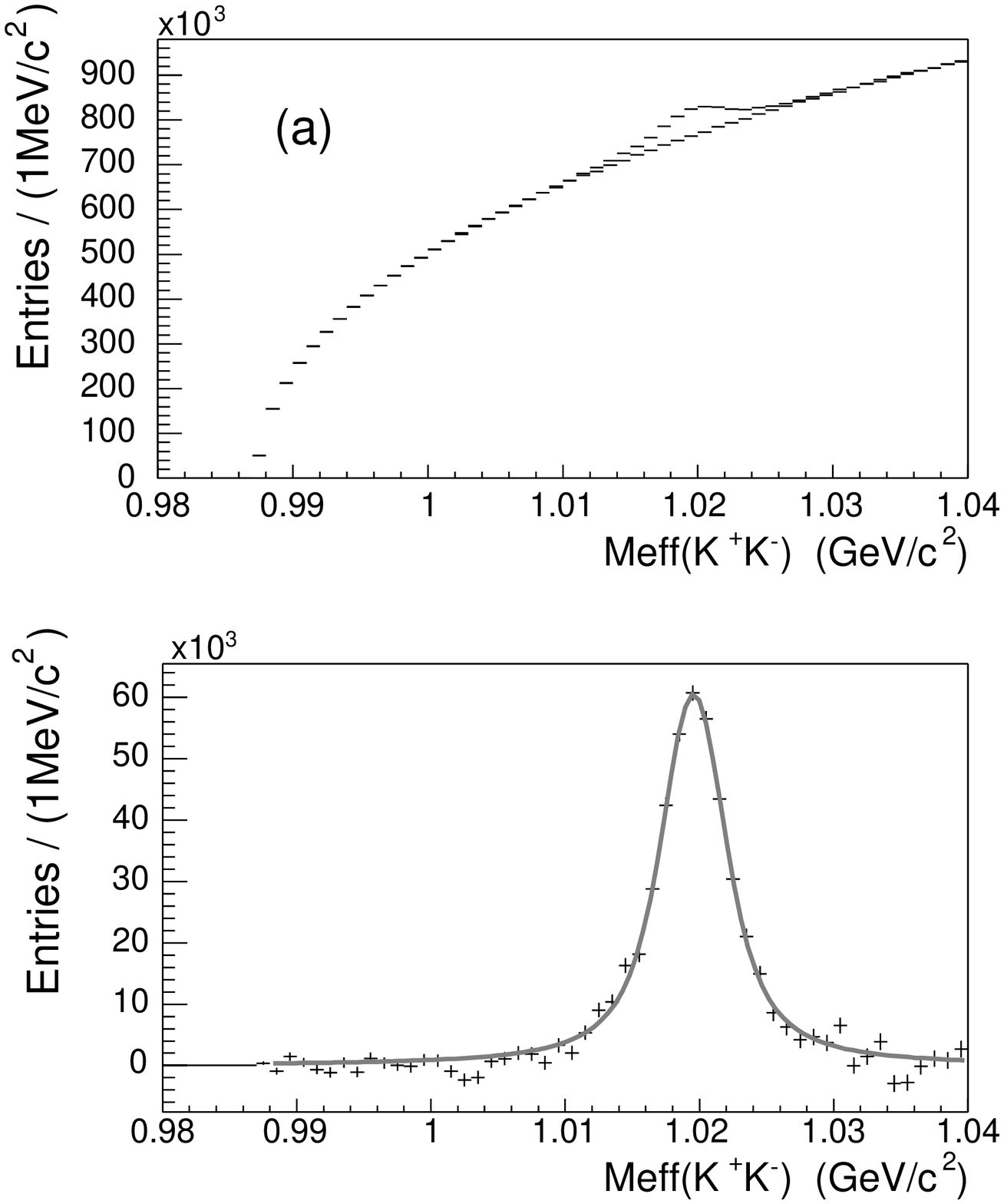}
\caption{Background subtracted invariant-mass spectra of K$^*$(890)$^0$
  (left panel) and $\phi(1020)$ (right panel) reconstructed in simulated 
  central \mbox{Pb--Pb} events.
\label{fig:resonances}}
\end{figure}

Resonances with lifetimes comparable to that of the QGP phase (such as 
$\rho^0$, K$^*$(890)$^0$, $\phi(1020)$) may 
change their properties (mass, width) when they are produced in the
dense medium. This may happen due to the final state interaction (which
can be sorted out by comparing the results obtained in the hadronic and
leptonic decay channels) or due to the partial chiral symmetry restoration.
Detection of  K$^*$(890)$^0$ and $\phi(1020)$ is especially important
because of the expected overall strangeness enhancement in heavy-ion
collisions~\cite{Rafelski:1982pu}.
    The ALICE detector will be able to register these resonances with
sufficient statistics and the mass resolution of the order of a few~MeV/$c^2$ 
(see Fig.~\ref{fig:resonances}).
   
   \subsection{Heavy-flavour QGP probes}
When traversing the dense matter created in nucleus-nucleus collisions,
the initially-produced hard partons lose energy mainly on account of
medium-induced gluon radiation. The heavy quarks
at intermediate \pt will lose less energy as compared with the light
quarks at the same momenta due to the `dead-cone' effect~
\cite{Dokshitzer:2001zm,Armesto:2003jh}.
The ratio of the nuclear modification factor for D (and B) mesons to the one
for the normal hadrons is thus
suggested to be sensitive to the mass dependence of in-medium parton
energy loss. The D mesons needed for these studies will be topologically 
reconstructed in central ALICE detectors, whereas B mesons can be detected
in semi-leptonic decay channels.

Precise detection of the open charm (open beauty) is also need for the
normalization of the quarkonia production. In nucleus-nucleus collisions, 
quarkonium suppression is expected to occur due to the Debye screening
in the deconfined matter. Increase of the energy density reached in the
collisions leads to break up of first the \mbox{$\psi^\prime$} and 
\mbox{$\chi_{\rm c}$}, and finally the  
\mbox{J\kern-0.05em /\kern-0.05em$\psi$}. Such a suppression pattern was
already observed at SPS~\cite{Alessandro:2004ap}. 
However, at higher energies (RHIC, LHC) the situation becomes more
complicated, because the charmonia can be regenerated
in the hot medium by recombination.
  
ALICE will be able to measure the charmonia and bottomonia production both
at mid-rapidity (in di-electron channels) and at forward rapidities (in
di-muon channels).  The statistics of registered \mbox{$\Upsilon$}'s
is expected to be enough for the suppression studies. In this case, due to much
higher mass, the influence of the regeneration will be negligible. 

   \subsection{High-\pt QGP probes}
High-\pt partons produced in initial hard parton-parton scatterings
fragment into jets. In the dense deconfined matter, prior to
hadronisation, the partons undergo significant energy losses. The evidence
of parton energy loss was observed at RICH as the suppression of high-\pt
particles~\cite{Adcox:2001jp,Adler:2002xw} and the suppression of back-to-back
correlations~\cite{Adler:2002tq}.

\begin{figure}
\centering
\epsfig{figure=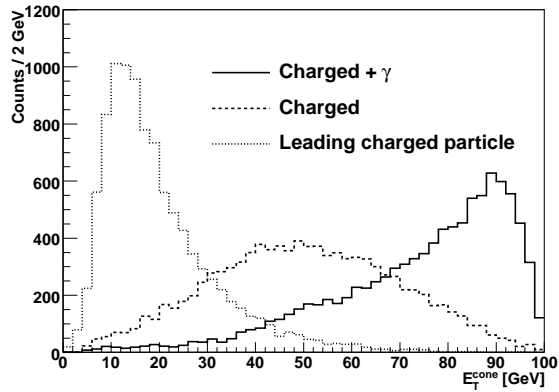,width=0.5\textwidth}
\caption{Spectra of reconstructed energy for simulated 100~GeV/$c$ jets.
The results for the charged leading particle, all charged particles,
and all charged particles and gammas included in the analysis are shown.
\label{fig:jetenergy}}
\end{figure}

    Due to harder \pt spectrum of jets at LHC, and ALICE's exceptional tracking
capabilities, the experiment will study the jet-jet kinematics with the 
precision unreachable in the leading-particle approach. Up to 50\% of the 
energy
of 100 GeV jet can be reconstructed with the ALICE detectors registering
the charged particles only. When ALICE is upgraded with the Electro-Magnetic
Calorimeter (EMCAL)~\cite{emcal}, the fraction of reconstructed energy will increase
even more (see Fig.~\ref{fig:jetenergy}). The EMCAL will also allow for
effective triggering on the jets.

For the first time, the expected statistics of reconstructed jets will be 
enough for differential studies. For example, the $k_{\rm t}$ spectrum of particles
belonging to a jet ($k_{\rm t}$ being the projection of particle momentum
on the plane perpendicular to the jet axis) is predicted to be 
significantly broader, when jets are produced in dense
medium~\cite{Salgado:2003rv}. The effect (especially the mean $k_{\rm t}$ and
the high-$k_{\rm t}$ tail) is enough stable with respect to the cuts
inevitably applied during the jet reconstruction.

Using the ALICE particle-identification capabilities, interesting 
observations can be also done with jets having identified particles,
in particular, the jets initiated by the heavy quarks (`dead cone'
related structures).

\section{Conclusions}
ALICE experiment at LHC is aiming to study the physics of strongly
interacting matter at extreme energy densities, where a new state of
matter, quark-gluon plasma, is expected to be reached.

The ALICE physics programme ranges from a precision measurements of the
bulk of matter created in heavy-ion collisions, with typical momenta below
0.5 GeV/$c$, to heavy quark physics, quarkonia spectroscopy and jet
measurements well above 100~GeV/$c$. Due to its excellent track,
vertex-finding and particle-identification capabilities, ALICE will be able
to study the properties of QGP by means of a whole set of different and
complementary observables.

\end{document}